# Spectromicroscopic measurement of surface and bulk band structure interplay in a disordered topological insulator

Erica Kotta,[1] Lin Miao,[1,2] Yishuai Xu,[1] S. Alexander Breitweiser,[1,3] Chris Jozwiak,[2]

Aaron Bostwick,[2] Eli Rotenberg,[2] Wenhan Zhang,[4] Weida Wu,[4] Takehito Suzuki,[5]

Joseph Checkelsky,[5] and L. Andrew Wray,[1]*

[1] Department of Physics, New York University, New York, New York 10003, USA
[2] Advanced Light Source, Lawrence Berkeley National Laboratory, Berkeley, CA 94720, USA
[3] Department of Physics and Astronomy, University of Pennsylvania, Philadelphia, PA 19104, USA
[4] Rutgers Department of Physics and Astronomy, Rutgers University, Piscataway New Jersey 08854, USA
[5] Massachusetts Institute of Technology, Department of Physics, Cambridge, MA, 02139, USA

* To whom correspondence should be addressed; E-mail: lawray@nyu.edu.

Topological insulators are bulk semiconductors that manifest in-gap massless Dirac surface states due to the topological bulk-boundary correspondence principle [1-3]. These surface states have been a subject of tremendous ongoing interest, due both to their intrinsic properties and to higher order emergence phenomena that can be achieved by manipulating the interface environment [4-11]. Here, ARPES spectromicroscopy and supplementary scanning tunneling microscopy (STM) are performed on the model topological insulator $Bi_2Se_3$ to investigate the interplay of crystallographic inhomogeneity with the topologically ordered bulk and surface band structure. Quantitative analysis methods are developed to obtain key spectroscopic information in spite of a limited dwell time on each measured point. Band energies are found to vary on the scale of 50 meV across the sample surface, enabling single-sample measurements that are analogous to a multi-sample doping series (termed a "binning series"). Focusing separately on the surface and bulk electrons reveals a nontrivial hybridization-like interplay between fluctuations in the surface and bulk state energetics.

The three-dimensional topological insulator (TI) state in $Bi_2Se_3$ was identified in 2008 [12,13] and features a large bulk band gap of ~0.3eV, spanned by a relatively ideal Dirac cone surface state. Though the existence of the surface state is protected by topology, the symmetries and band dispersion of surface state electrons can be strongly influenced by crystallographic defects, potentially including the emergence of new quasiparticles [7-11,14-18]. Bulk grown samples also feature inhomogeneity on the 1-1000 μm scale, and the diversity of chemical potential, lattice strain, and disorder at the micron scale encodes tremendous information about the electronic system that would be difficult to reproduce through the deliberate control of doping, strain, or other experimental parameters. Thus a major purpose of this investigation is to explore rigorous procedures for utilizing this intrinsic inhomogeneity to understand the variability of Dirac electronic structures (see summary of analysis procedures in Methods).

The central region of the sample surface, exposed by *in situ* cleavage, was mapped in 961 individual ARPES images covering a 31 x 31 grid with uniform 10 μm spacing. A dwell time of just 2s/point was used to avoid surface aging (band bending [19]) from the highly focused incident photon beam (see characterization in Supplemental Fig. S1). Binning all 961 measurements results in a high-statistics Dirac cone image that is equivalent to a conventional ARPES measurement performed with a .3 x .3 mm$^2$ beam spot (Fig. 1a). An example of a single-point 2s spectrum is shown in Fig. 1b, with the otherwise sparse data broadened for visibility. A V-shaped upper Dirac cone can be immediately identified in most single-point ARPES spectra, but the statistics are inadequate for a close analysis without first leveraging information from the rest of the surface.

To map out the variation in the surface state electronic structure across the sample surface, a small window around the Dirac point of the spatially averaged ARPES image (boxed in red in Figure 1a) is used as a template for normalized cross-correlation with each single-scan spectrum (see Supplementary Note 3). The Dirac point coordinate $\{E_D, k_D\}$ is identified from a sharp peak in the cross-correlation map (Fig. 1c), yielding surface maps shown in Fig. 2a,c. As the Dirac point is constrained by symmetry to always be located in the Brillouin zone center, shifts in $k_D$ actually represent changes in the orientation of the surface, which changes the angle of surface-normal (k=0) photoemission.

Shifting each ARPES image along the momentum axis ($k_x$-axis) to set $k_D$=0 therefore allows for a more meaningful binning of the images and makes it feasible to fit other quantities such as the *bulk* band energies. For example, by examining a constant-momentum energy dispersion curve (EDC) evaluated at $k_D$, one can fit the onset of intensity with a step function to determine the bulk conduction band minimum energy (map in Fig. 2c). Beyond these maps of binding energy, evaluating scattering intensity in a small $\{E,k\}$ range centered around the cross-correlation maximum $\{E_D, k_D\}$ point offers a quick diagnostic tool to identify anomalous regions in which the surface state less closely resembles an ideal Dirac cone (Fig. 2d).

The resulting maps of these four properties reveal very different forms of spatial structure on the sample surface. The map of Dirac point momentum ($k_D$, Fig. 2c) shows a gently warped surface with continuous gradations and few sharp boundaries. Two triangular regions, shaded in gray, show a much rougher topography and are excluded from further analysis (see Supplementary Fig. 3). The Dirac point energy map (Fig. 2a) shows a significant ~50 meV distribution, with approximately half (48%) of the surface lying within 10 meV of the average $E_D$~440 meV binding energy. Higher binding energies are found in several ~50 μm diameter islands, and a strip with lower binding energy runs along the bottom edge of the sample.

The bulk band energy map (Fig. 2b) reveals fluctuations on approximately the same ~50 meV scale as the surface Dirac point, and a similar horizontal trough feature with shallow binding energies is recognizable near the bottom edge. Other than this, surface regions look qualitatively different from the Dirac point energy map, and regions with deeper bulk binding energy show limited correlation between the two maps, as we will see in Figures 3-4. The Dirac point intensity map (Fig. 2d) is smoothest of all, showing a flat surface transected by three dark stripe-like features running diagonally downwards from left to right. These are most likely caused by buckling of the sample along the y-axis, which causes the measurement to slightly miss the Dirac point in 2D momentum space (see Supplementary Fig. 3b). Faint shadows of these 3 features are also observable in the Dirac point momentum ($k_D$) map in Fig. 2c, however only the feature in the lower left corresponds with a clear feature in the bulk and surface energy maps in Fig. 2a-b.

Supplementary STM measurements were performed to investigate the chemical origin of these variations in binding energy and confirm that point defect populations vary greatly across the surface. Four ≥ 100 x 100 $nm^2$ surface regions separated by ~0.5 mm were measured on a sample from the same growth batch, revealing that charge-donating point defects within 1nm of the surface are primarily intercalated excess Se ($Se_I$) and Se replacement sites ($Bi_{Se}$) associated with excess Bi. The densities of these two defect types were strongly anticorrelated, representing inhomogeneity in the relative local density of Bi and Se (see Supplementary Table 1). On average, there was a stoichiometric excess of 0.0014 Bi atoms per 5-atom unit cell, with a standard deviation of 0.0004 (±30%) for different sampled regions, sufficient to greatly influence the chemical potential and surface state energetics [7,8].

Having corrected for the surface curvature, the large fluctuations we observe in bulk and surface binding energies can now be used to organize the data in a way that resembles a doping series, but which we will term a "binning series". Figure 3a1-a4 shows composite spectra obtained from binning measurements with similar Dirac point binding energy. Spectra are labeled with the averaged center-of-mass Dirac point energy of the measurements binned within them, and the k=0 EDC of each spectrum is shown to the right (Fig. 3b). A similar series binning from the bulk energy map is displayed below in Fig. 3c-d. The single-point measurements used for these binning series images are indicated via a color code in histograms (Fig. 3e-f) and on the full surface maps (Fig. 3g-h).

The Dirac point series EDCs (Fig. 3b) show a peak at the Dirac point energy that nicely tracks the average bin energy. However, the onset of the bulk conduction band at $E_B$~0.23 eV binding energy shows surprisingly little dependence on the Dirac point energy, suggesting that different factors may be responsible for the fluctuations in bulk and surface bands. Larger bin widths were used in the bulk binning series (see Fig. 3f histogram), for which we will briefly discuss bias effects intrinsic to this presentation of the data. Each image in the bulk series (Fig. 4c1-c3) is labeled with the energy center of mass, and these numbers span a 32 meV range. However, these numbers are biased, due to the points being drawn with some error from a peaked distribution. A conservative estimate of this bias can be obtained by characterizing the error in the fitting procedure, then evaluating the convolution of this error with the distribution from which the measurement is sampled (see Supplementary Note 4). This results in bias-corrected bin energies with just a 20 meV distribution (0.227, 0.240, and 0.247 eV for Fig. 3c1-c3, respectively). Less error is expected in the Dirac point series, due to the smaller bin sizes and slightly superior fitting error.

To better understand the interplay between bulk and surface energetics, it is useful to resolve how surface state energies in different parts of momentum space change in response to this 20 meV shift of the bulk band. The change in intensity at the leading edge of the surface state can be used to estimate this shift, as $\Delta E_S(k)=[I(E,k)_{4c3} - I(E,k)_{4c1}] \div [\partial I(E,k)_{avg}/\partial E]$. Here, $I(E,k)_\alpha$ is an ARPES image indexed by $\alpha$, and 'avg' indicates an average over the spectra in panels 3c1-c3. The results are summarized in Fig. 4a, where the length of the blue arrows is set proportional to the magnitude of the surface state energy shift at that momentum. An alternative analysis method is presented in the Supplementary Figure S6 and shows roughly equivalent results.

The picture obtained is intriguingly non-uniform. Surface energy shifts are relatively large (>10 meV from Fig. 4a) for $|k| > 0.007$ Å$^{-1}$ and $|k| < 0.003$ Å$^{-1}$, corresponding to a ~55% correlation with the 20-meV bulk band shift. At the intermediate values $0.003$ Å$^{-1}$ $<$ $|k|$ $<$ $0.007$ Å$^{-1}$ where the surface state falls in the middle of the bulk band gap, the shifts dip beneath 7 meV (Fig. 4a), indicating a smaller ~35% correlation. The observation that the surface state moves in tandem with the bulk band near the Fermi level and Dirac point while being less correlated at the intermediate momenta can be partially understood from the fact that the surface state is intrinsically hybridized with the bulk continuum, inducing a greater correlation with the bulk energetics. Near the Fermi level, the surface state runs along the external contour of the bulk conduction band with a near-parallel trajectory [6], and shifts strongly and uniformly in tandem with the change in the bulk binding energy. At larger binding energies near the Dirac point, strong hybridization is again expected from proximity to the bulk *valence* band. In Fig. 4b, a tight-binding model for $Bi_2Se_3$ (see Supplementary Note 7) is used to investigate the numerical derivative of surface state energy as the bulk energy is shifted ($dE_S(k)/dE_B$), with mid-gap surface state energy held fixed to enhance contrast. Bulk energy was controlled by increasing the binding energy of all orbitals beneath the outermost quintuple layer, resulting in a downward distortion of portions of the surface state that were energetically close to the bulk bands. A maximum amplitude of $dE_S(k)/dE_B$ = 18% is seen at the Fermi level, corresponding to the expected difference between the surface-bulk energy correlation at the Fermi level and at the middle of the bulk band gap. This value is remarkably close to the ~20% difference seen within the experimental data, suggesting that a hybridization-based interpretation can plausibly account for the evolution of the surface state across the bulk binning series.

A further prediction one can draw within this picture is that when the bulk bands have large binding energy and are thus 'pulled away' from the surface Dirac point, bulk-surface energy correlation at the Dirac point should become smaller. Figure 4c shows a histogram of the surface Dirac point ($E_S$) versus bulk ($E_B$) energies, with the center of mass trend traced as a function of bulk energy (dashed purple line). Center of mass coordinates are plotted for each $E_B$ bin with more than 10 measurements, and use a larger step near the edges of the plot to maintain a minimum of 10 measurements per point. This trend curve has a significant slope at the center of the bulk binning series ($E_B$=0.24 eV), which increases as the bulk band is brought to shallower binding energy (to $E_B$~0.19 eV), and flattens out as it is brought to deeper binding energies, matching expectations for a hybridized picture (solid blue line). A kink in the trend curve at $E_B\sim0.18$ eV is also present in the simulation, where it occurs due to the transition of the Dirac point from a true in-gap surface state to a bulk-degenerate 'resonance state'.

Taken as a whole, the surface maps obtained from spectromicroscopy on $Bi_2Se_3$ reveal great complexity on the 10-100 μm scale, with very different sets of structural features visible in the surface curvature, Dirac point intensity, and band energy maps. Fluctuations in band energetics have a sizable ~50 meV scale, and can be partially understood from STM topography maps, which reveal very different defect population distributions in surface locations separated on the micron scale, stemming from a large ±30% variation in the amount of excess Bi. Tracking specific band features associated with the surface and bulk of the material reveals that the spectromicroscopy data set contains a remarkably diverse range of energetic scenarios distributed over a *multi-dimensional* parameter space. To make use of this, we develop a methodology for analyzing data in a "binning series" that reveals changes in band dispersion as a function of a single parameter. Organizing such a binning series in terms of bulk energetics shows a pattern of surface-bulk correlations that can be interpreted from a hybridization picture, in which bulk and surface states strongly repel one another when they come within ~50 meV of intersecting. This analysis illustrates the potential of ARPES spectromicroscopy as a tool to investigate the interplay of electronic states via hybridization and strong correlation effects, with a richness that cannot be captured from a single-dimensional doping series.

## Methods

### *Sample growth, characterization and measurement*

The sample was grown following standard procedures of [20,21], but with an abbreviated annealing process of 24 hours (compared to the standard ~3 days) to achieve a single-phase sample that retains a relatively high defect density. The sample was cleaved and measured *in situ* at low temperature (T~20K) and ultra-high vacuum (<5×10$^{11}$ Torr) at the microARPES endstation of the Advanced Light Source MAESTRO 7.0.2 Beamline. Measurements were performed with a photon energy of 110 eV, and photoelectrons were passed through a Scienta

R4000 electron analyzer for an energy and momentum resolution better than 30 meV and 0.01 $Å^{-1}$, respectively. The illuminated beam spot was smaller than δ~<15 μm in diameter. All measurements were performed along the Γ-K momentum axis of the hexagonal surface Brillouin zone, along which the Dirac cone appears as a V-shaped contour centered at the Γ point. The measurements made use of π-polarization (predominantly out-of-plane), which maximizes the overall ARPES signal, but has been suggested to give greater spectroscopic weight to surface regions with a lower defect density [8].

*Summary of analysis procedures*

The analysis procedures demonstrated here can be adopted independently, but also constitute a sequentially cumulative recipe, using the following procedural order:

1. **Normalization:** Represent ARPES intensity in terms of number of the number of recorded photoelectrons. A simple intensity calibration mask can suffice. This step is the prerequisite for simulating analysis/fitting approaches to achieve quantitative error bars (see Supplementary Note 3).

2. **Cross-correlation:** Track the energy and momentum coordinates of the Dirac point. Error bars can be obtained by simulating the cross correlation procedure, or from the standard deviation of regional fluctuations. This step is the prerequisite for applying momentum-axis corrections

3. **Fit/Extract other band properties:** Correct all data with the momentum shifts obtained in step 2, and then apply additional measurements/fits to this data set. Examples in the paper include Dirac point intensity mapping, and bulk band energy fitting. A quantitative error estimate can be obtained from simulating the measurement based on a normalized template.

4. **Create composite spectra:** Histogram the measurements as a function of individual quantities extracted in steps 2-3 (e.g. Dirac point energy), and create composite spectra to evaluate how band structure varies as a function of the quantity. If the histogram distribution is sharply peaked (not flat), then it is vital to correct for sampling error as discussed in Supplementary Note 3.

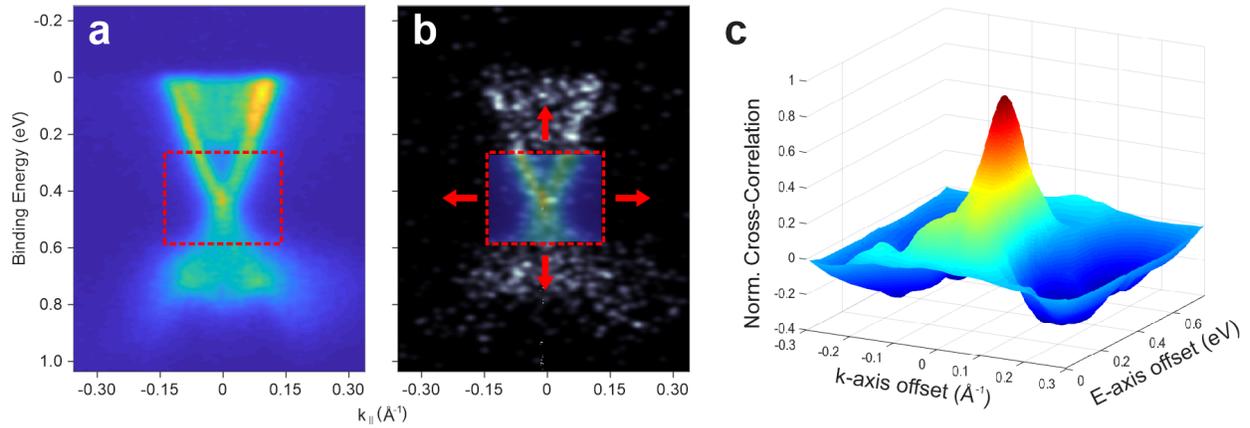

**Figure 1: Cross-correlation for ARPES analysis. a,** A composite spectrum obtained by summing all data from a 31x31 point spectromicroscopy map of the sample surface. A red square enclosing the surface state Dirac point indicates the data window used as a template for cross correlation. **b,** An individual scan is overlaid with the template cut-out from panel (a). Gaussian broadening with a half-width of 13 meV and 0.015 $\text{Å}^{-1}$ is applied for visibility. **c,** The energy- and momentum-space location of the surface Dirac point can be seen from a sharp peak in the normalized cross-correlation of the composite template with the individual scan data from panel (b).

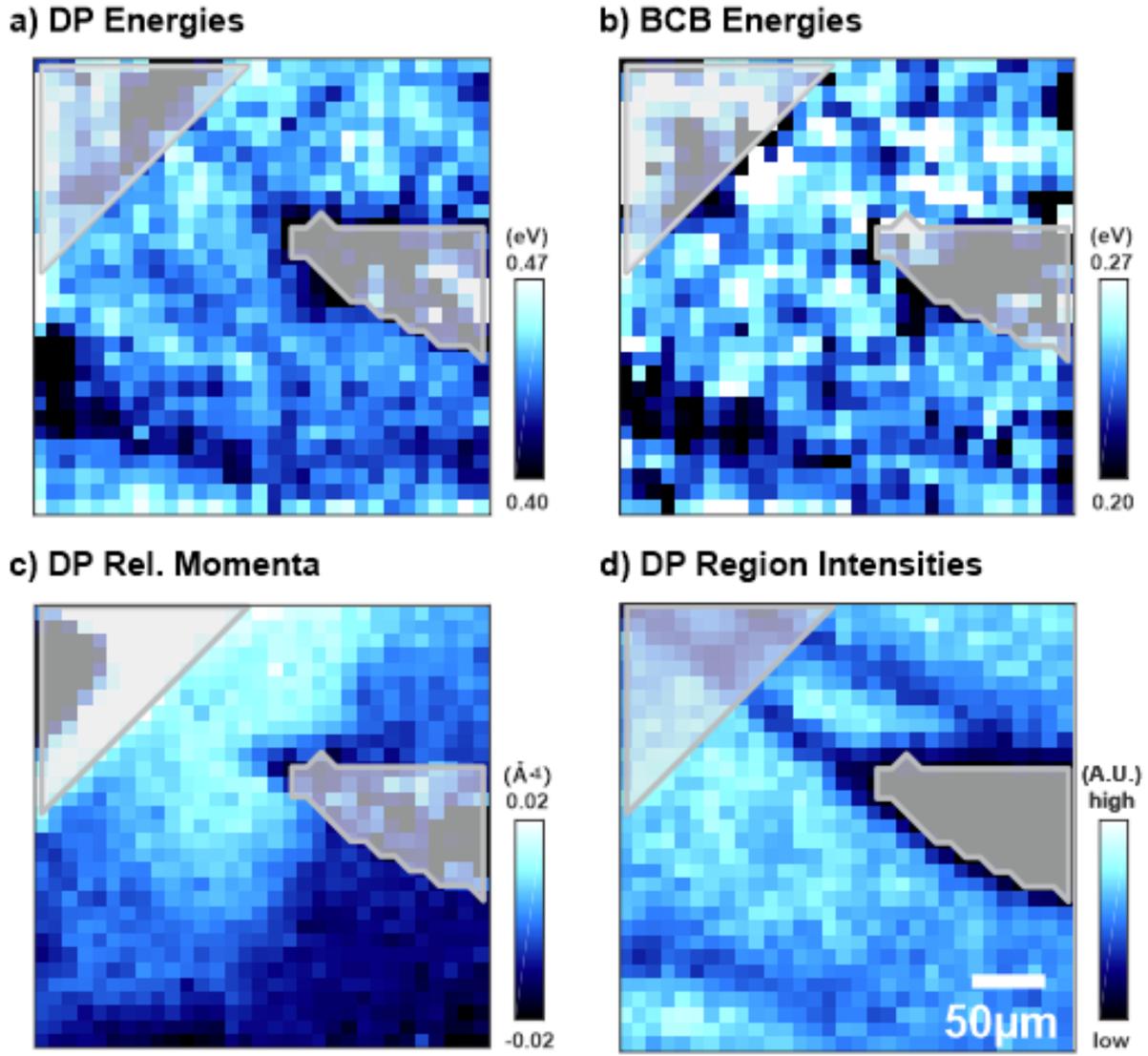

**Figure 2: Real space maps of electronic band structure. a**, The surface state Dirac point binding energy. **b,** The bulk conduction band energy minimum, with data smoothed by adding 25% of the nearest neighbor pixel intensities. Error bars for panels (a-b) are <~10 meV (see Supplementary Note 3). **c**, The relative shift of the Dirac point on the $K_x$-axis of momentum space, representing the direction of the surface-normal cleavage axis. **d**, Spectral intensity at the Dirac point. The gray shaded regions do not contain a Dirac-cone-like spectral feature and are excluded from detailed analysis.

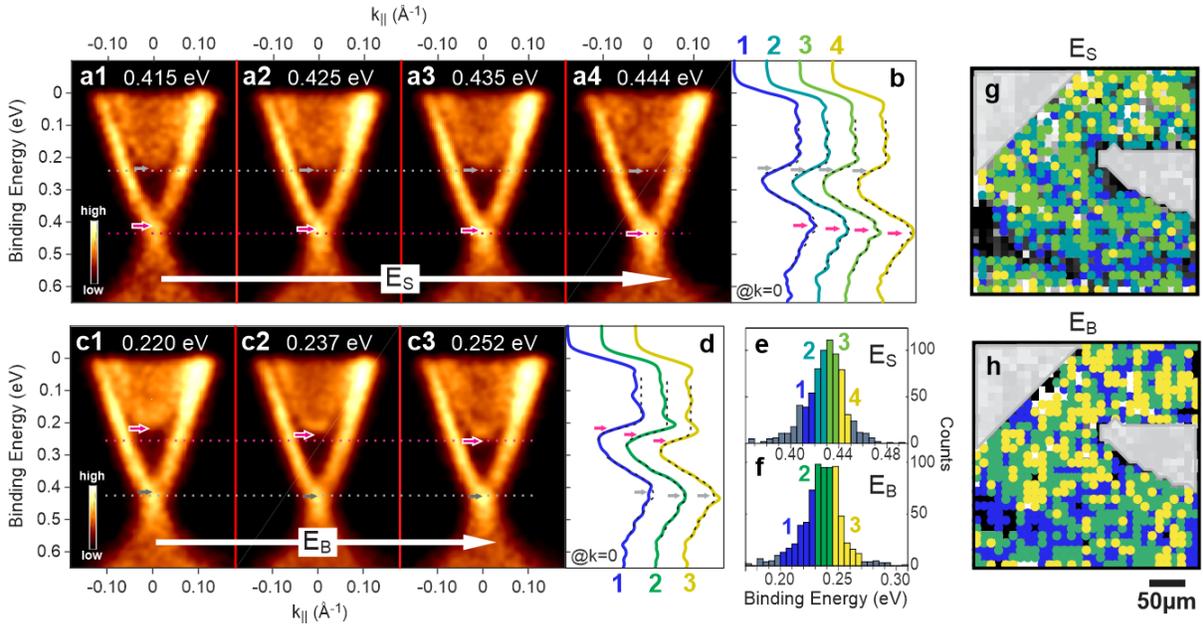

**Figure 3: Bulk- and surface-ordered binning series. a1-4,** The ARPES spectra with Dirac point energies within ±5 meV of the labeled energy are binned to create composite images. **b,** Constant momentum profiles of spectral intensity (EDCs) in the Brillouin zone center taken from the spectra a1-4. Grey arrows indicate the bulk band onset energy, and magenta arrows indicate the Dirac point energy. Templates used for fitting are shown as black dashed lines. **c1-3,** ARPES spectra with bulk binding energies within wider, variable-width ranges are binned together and labeled with the center-of-mass energies of the corresponding bin. **d,** Brillouin zone center EDCs for c1-3. **e-f,** Histograms of the observed (e) Dirac point and (f) bulk energies are color-coded to indicate the regions used for binning series panels. **g-h,** Maps of surface and bulk energies are color-coded as defined on the panel (e-f) histograms to indicate their incorporation within the binning series sets.

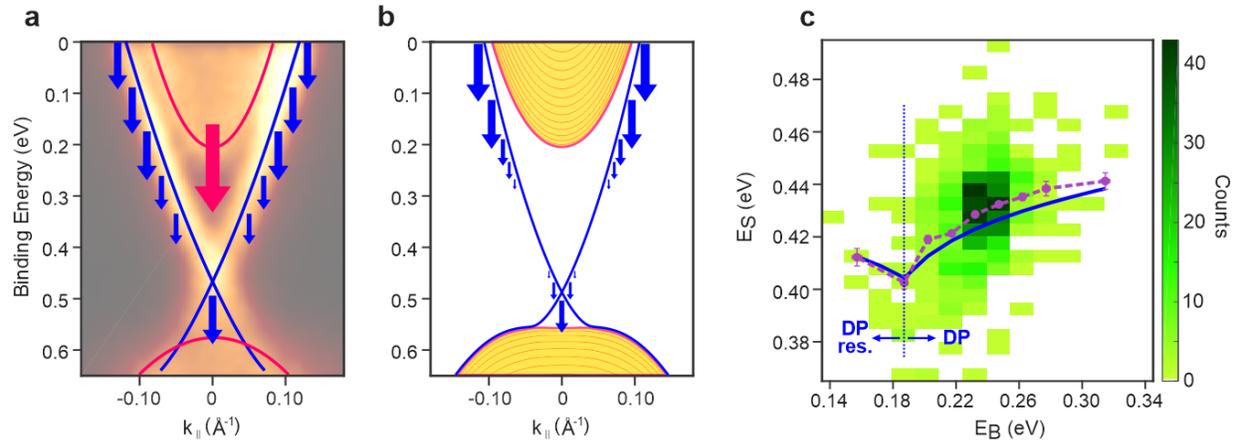

**Figure 4: Interplay of bulk and surface electronic states. a,** A summary image showing how different portions of the surface state shift in energy across the bulk band binning series. The length and thickness of the arrows are scaled linearly with the magnitude of the energy shift of the surface band at that momentum across the bulk binning series, with an amplitude of $\Delta E_S$=11 meV attributed at the Fermi level ($\Delta E_S/\Delta E_B \sim$ 55% = 11/20). **b,** A simulation of the repulsion between surface and bulk bands, with the mid-gap surface state (at k=0.05Å$^{-1}$) set to a fixed binding energy. Arrows show the fractional energy shift in response to the change in bulk energies ($dE_S/dE_B$ = 18% at the Fermi level) when a uniform potential is applied to all orbitals beneath the outermost quintuple layer. **c,** A 2D histogram of the observed Dirac point and bulk conduction band energies is overlaid with a purple trend curve showing the $E_B$-resolved center of mass (see text). Error bars represent propagated single-point measurement error. A simulation of Dirac point (DP) energy versus bulk energy is plotted in blue. At $E_B<\sim$0.18 eV the modeled spectral feature leaves the band gap and becomes a bulk-degenerate resonance state (DP res).

*Supplementary information for*

# Spectromicroscopic measurement of surface and bulk band structure interplay in a disordered topological insulator

Erica Kotta,[1] Lin Miao,[1,2] Yishuai Xu,[1] S. Alexander Breitweiser,[1,3] Chris Jozwiak,[2]

Aaron Bostwick,[2] Eli Rotenberg,[2] Wenhan Zhang,[4] Weida Wu,[4] Takehito Suzuki,[5]

Joseph Checkelsky,[5] and L. Andrew Wray,[1]*

[1] Department of Physics, New York University, New York, New York 10003, USA
[2] Advanced Light Source, Lawrence Berkeley National Laboratory, Berkeley, CA 94720, USA
[3] Department of Physics and Astronomy, University of Pennsylvania, Philadelphia, PA 19104, USA
[4] Rutgers Department of Physics and Astronomy, Rutgers University, Piscataway New Jersey 08854, USA
[5] Massachusetts Institute of Technology, Department of Physics, Cambridge, MA, 02139, USA

* To whom correspondence should be addressed; E-mail: lawray@nyu.edu.

**Supplementary Note 1: Sample aging and measurement time**

Characterization of a surface point (from the same sample, but outside the region analyzed and mapped for this paper) revealed that the Dirac point shifted by roughly 100 meV in 5 minutes of beam exposure (Supp. Fig. 1). This appears to suggest that aging will modify the surface energetics by just $\delta E < \sim 0.5$ meV, however this linear extrapolation will greatly underestimate the rate of change in the first seconds of exposure, and it is safe to assume that the actual change is of a larger scale (possibly ~10 meV).

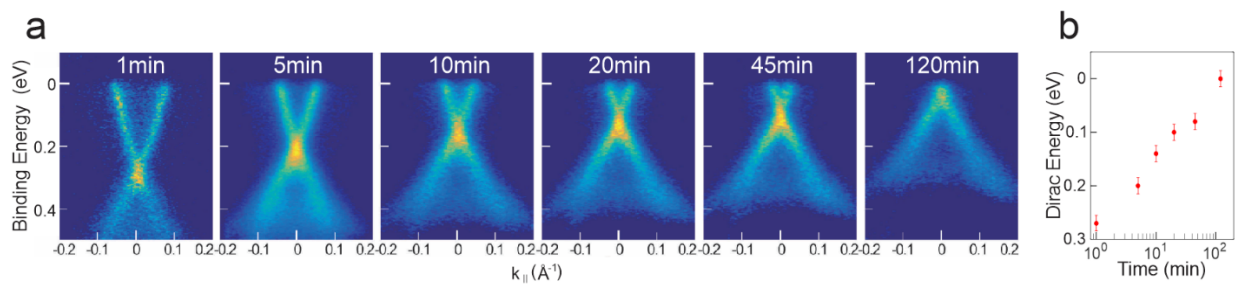

**Supplementary Figure 1: Aging effect.** Spectra taken at the same beam line, but with a larger beam spot of 30x40 um$^2$ and a slightly higher photon energy of 117 eV (as opposed to 15x15 um$^2$ beam spot and 110 eV for the measurements analyzed in this paper), over progressively longer dwell times. This dramatic shift in the Dirac point was the motivation for choosing a smaller beam spot size (15x15 um$^2$) and using extremely short dwell times (~2 seconds per pixel), although requiring more in terms of data analysis techniques to work around the decreased signal-to-noise. This trend has the opposite sign to that observed in [19].

**Supplementary Note 2: Normalizing the detector intensity unit to events count**

Pixel sensitivity on a CCD camera screen tend to gradually vary over time and years of near-continuous measurements, so steps were initially taken to renormalize the pixel intensity of the raw ARPES images. First, areas far outside the Dirac cone, where events were sparse and due primarily to noise, were analyzed. Blobs—areas of connected activated pixels—were counted as individual events, and pixel intensities were rescaled such that the total pixel intensity for an average single blob summed to 1 (a.u.). Supplementary Figure 2 shows the summary of total pixel intensity in each of the four boxed regions in Supp. Fig. 2a over all 961 ARPES images, showing that the pixel intensities follow a Poisson distribution, consistent with random events. This allowed henceforth the direct interpretation of pixel intensity to photoelectron detection event.

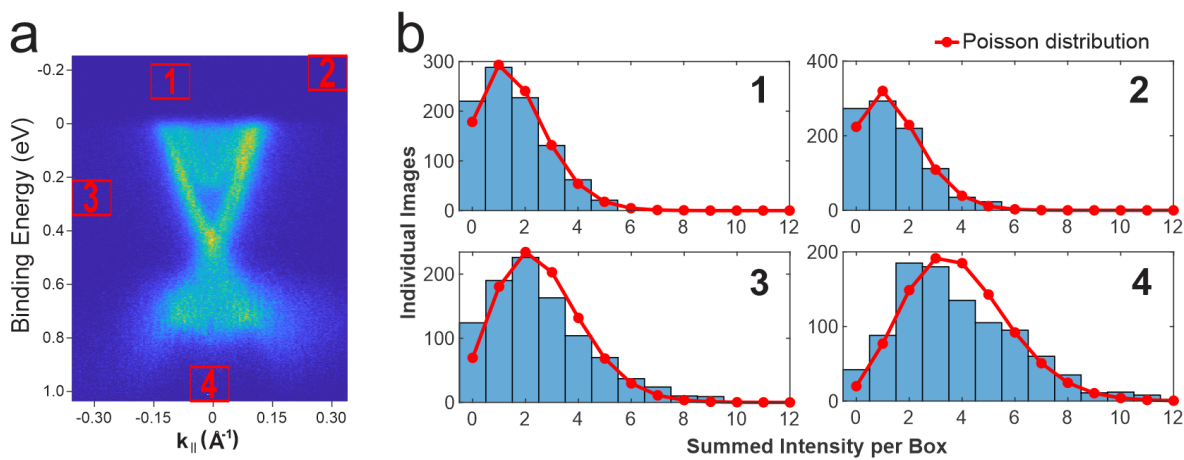

**Supplementary Figure 2: Poisson analysis.** Blobs in the sparse outside regions were analyzed and assumed as representing single events, then used to renormalize pixel intensities across the energy axis. For each box position 1-4, the total pixel intensities were summed over each individual ARPES image, and their distributions shown as bar plots. Poisson distributions of their respective averages are superimposed in red, showing a close qualitative comparison.

**Supplementary Note 3: Cross-correlation to track the Dirac point**

The normalized cross-correlation coefficients, performed on each individual ARPES image (I) using the full binned image window as a template (T), was calculated using the equation below:

$$X(E,k) = \frac{\sum_{E',k'}[I(E',k') - \bar{I}(E,k)] * [T(E'-E, k'-k) - \bar{T}]}{\sqrt{\sum_{E',k'}[I(E',k') - \bar{I}(E,k)]^2 * \sum_{E',k'}[T(E'-E, k'-k) - \bar{T}]^2}} \qquad (1)$$

At each template position (E,k), the sum over (E',k') is run over the range included underneath the template, $\bar{I}(E, k)$ is the average pixel intensity under the template at the position (E,k), and $\bar{T}$ is the average pixel intensity of the template.

The Dirac point Region Intensity Map was produced by taking the sum of the pixel intensities (in a fixed pixel range across all individual ARPES images) broadly encompassing the surface bands from slightly below the Dirac point and up to approximately 200 meV above it. While small deviations from the average region intensity value could speak either to the statistical quality of that particular image or to slight fluctuations off the $k_\perp = 0$ axis, those map points corresponding to excessively low (less than 1.5 standard deviations below the mean) region intensity values correspond to a fluctuation so far off the $k_\perp = 0$ axis as to be missing the Dirac point entirely. The identified images were omitted from all further analysis (Figures 3-4 and related discussion). Furthermore, upon looking at the results of the Dirac point relative momenta map (Fig. 2c), the upper left corner shows a region of highly anomalous structure; the ARPES images from this area were also omitted from further analysis/discussion.

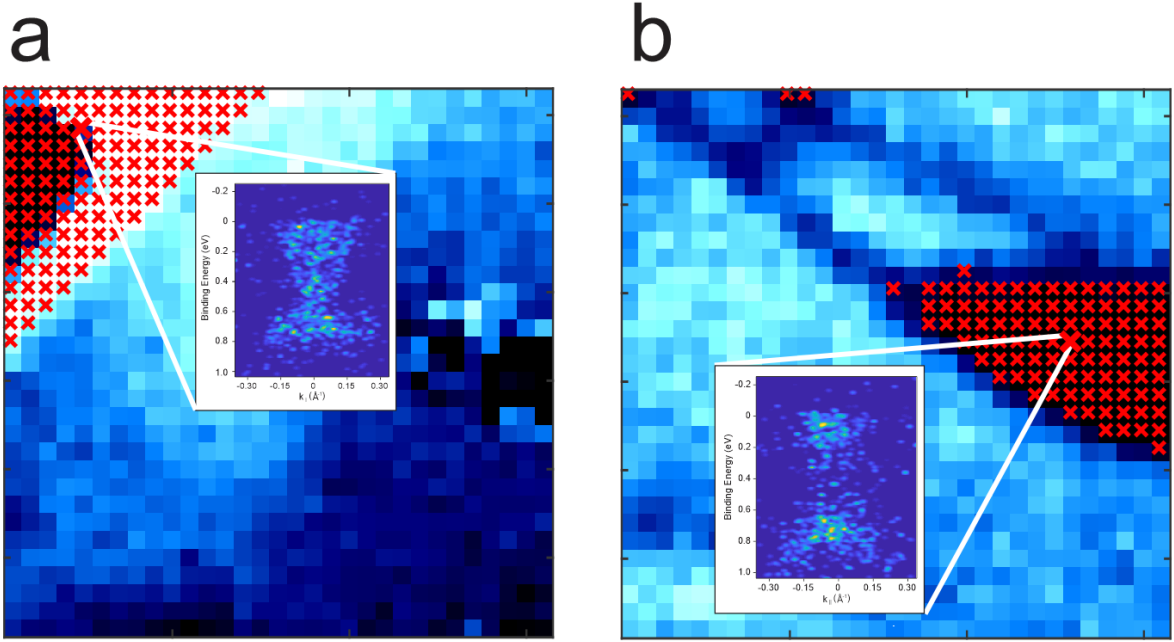

**Supplementary Figure 3: Omitted individual images.** **a,** Manually selected region from the *Dirac point* Relative Momenta map for individual ARPES images to omit from further spectra analysis. **b,** Those images with a Dirac point region intensity below a certain threshold were also omitted.

**Supplementary Note 4: Quantitative error analysis**

**Dirac point energies:** Examining a few flat regions taken from the Dirac point energy map (highlighted boxes in Supp. Fig. 4a), the standard deviation in Dirac point energy was calculated to be roughly 9±2 meV. To address concerns regarding sample bias of our relatively small and deliberately placed sample selections, we performed similar numerical analysis on same-sized regions of randomly distributed numbers and found the upper bound of the standard deviation to be 20.8%; this justifies our sample size and is consistent with our observation of the larger-scale structure apparent in the Dirac point energy map.

Since the energy bins used to integrate the spectra of Figure 3a were both relatively narrow and of fixed width (10 meV), the center Dirac point energies for Fig. 3a1-4 were approximated as simply the center of mass of the two contained 5-meV bins.

**Bulk conduction band minima energies:** Bulk energetics were extracted by taking an energy dispersion curve (EDC) at the Dirac point momentum coordinate $k_D$ and fitting a step function to find the BCB minimum. The lack of a sharp band-like contour for bulk electrons is due to the coincidence of the photoelectron escape depth with the surface state skin depth [22], resulting in the inability to nicely resolve z-axis momentum.  To quantify error in the bulk conduction band map, first an EDC @ k=0 was taken of the spectrum in Fig. 3c2. Then new EDC curves were simulated by selecting random numbers from a Poisson distribution about the original curve for each point along the energy axis, and the fit method performed on the simulated curve. Analyzing the results over many iterations, the method shows a standard deviation of 10 meV (see Supp. Fig. 4b-c).

The simulated distribution was then used as a representative sampling distribution and used to deconvolute the data distribution of the BCB energy fits. The deconvolution width parameter is merely an estimate, but it has a rough upper bound associated with the appearance of large negative-intensity artifacts in the deconvoluted spectrum. In the present case, this is approximately equivalent to a fitted upper bound based on the sharpness of features in the spectrum. Based on this, we estimate a maximal broadening factor from sampling of 7 meV. The sampling distribution $G$ (centered at 0 and normalized to 1) and the deconvoluted BCB distribution $I_0$ were then used to calculate the corrected center of mass energies of each Fig. 3c spectrum BCB energy discussed in the main text, using Equation 2 below:

$$< E > = \frac{1}{Z} \sum_{E_1 E_2} E_2 G(E_1 - E_2) I_0(E_2), \tag{2}$$

where

$$Z = \sum_{E_1 E_2} G(E_1 - E_2) I_0(E_2) \tag{3}$$

is the summed intensity of the region. Here, $E_1$ runs over all energy bin values, and $E_2$ runs only over energy (midpoint) values of the underlying bins that were included in the spectrum of consideration.

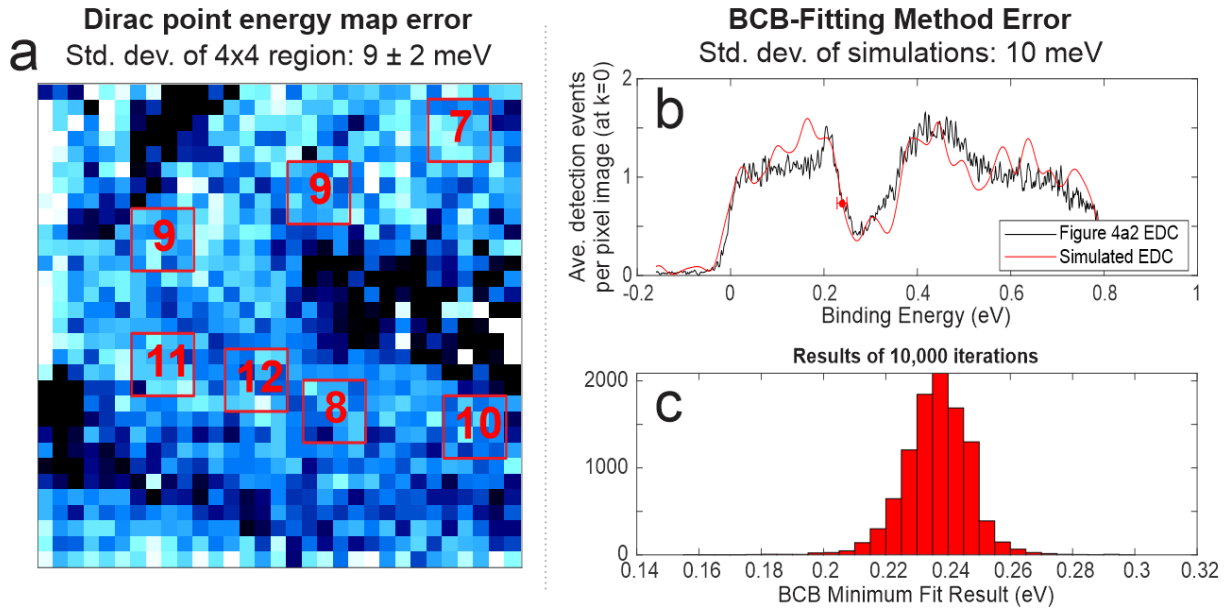

**Supplementary Figure 4: Quantified error for Dirac point energy and bulk conduction band minima maps. a,** The standard deviation was calculated for selected 4x4 pixel regions selected from relatively flat areas of the Dirac point energy map (labeled in units of meV). **b,** EDC curves were simulated from Poisson distributions about the EDC curve taken from Fig. 3c2. **c,** After many iterations, the standard deviation of the simulation results was 10 meV.

**Supplementary Note 5: STM images of point defects**

The following Supplementary Figure 5 shows STM images taken from a Bi$_2$Se$_3$ sample from the same batch as the one studied in this paper. Images L1-4 were taken at least 0.5 mm apart from each other. Three main defects were identified: intercalated Se above the top layer (Se$_i$ (0.5A), red circles) and in between the first and second QL (Se$_i$ (5.5A), red squares); Bi-Se antisites in the top layer (Bi$_{Se}$ (1A), blue circles) and in the fifth or sixth layers (Bi$_{Se}$ (5B or 6C), blue squares); and Se vacancies in the top layer (V$_{Se}$ (1A), green circles) and within the third or fifth layer (V$_{Se}$ (3C or 5B), green squares).

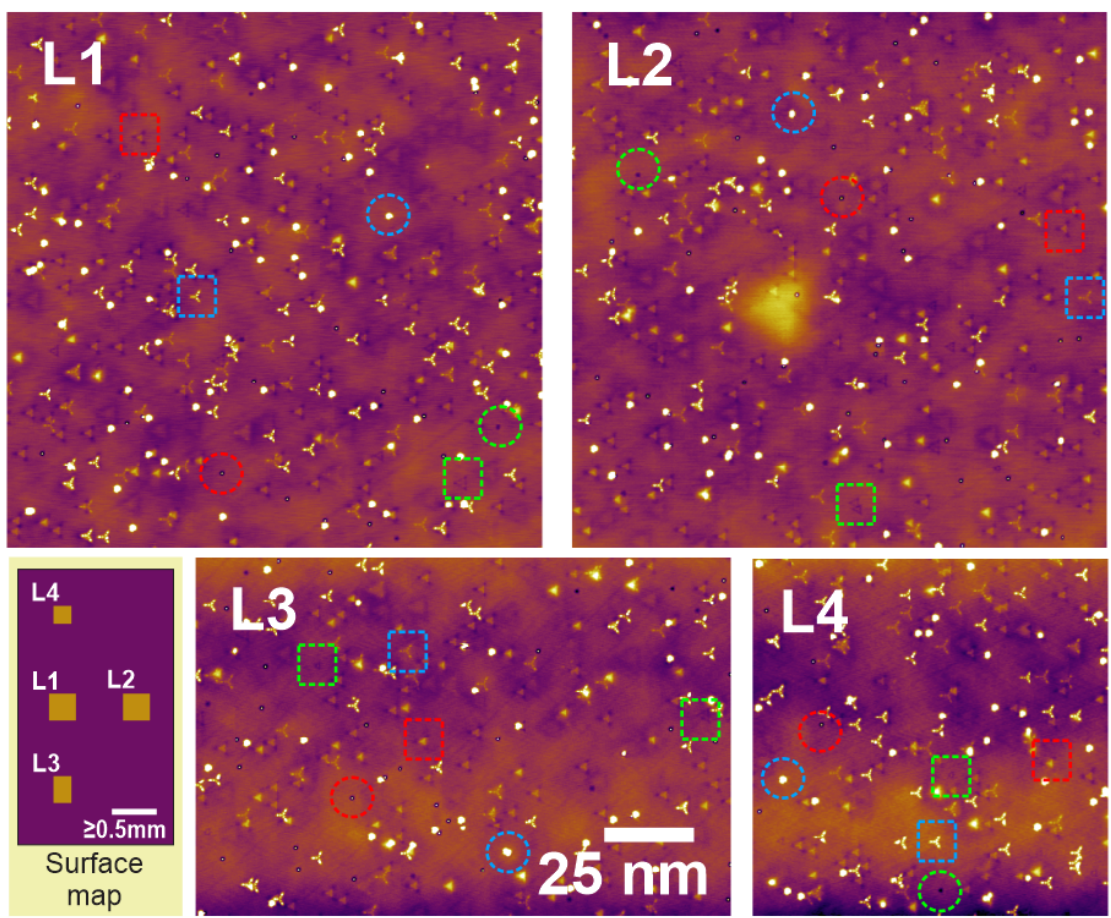

**Supplementary Figure 5: STM images of defects on Bi$_2$Se$_3$.** Images are taken from four different locations (L1-4) along a sample surface, with the three main types of point defects indicated: Se$_i$ (0.5A), Se$_i$ (5.5A) in red circles, squares; Bi$_{Se}$ (1A), Bi$_{Se}$ (5B/6C) in blue circles, squares; V$_{Se}$ (1A) and V$_{Se}$ (3C/5B) in green circles, squares. L1-4 images to same scale. Approximate locations L1-4 are shown in the surface map in the lower left (rotated 90°CCW relative to individual images; not to scale), and each pair of locations were at least 0.5mm apart.

The defects were counted and plugged into the following equation to estimate the stoichiometric excess Bi of the sample:

$$Excess\ Bi = -\frac{2}{3}Se_i^{0.5A} - \frac{1}{3}Se_i^{5.5A} + \frac{5}{3}Bi_{Se}^{all} + \frac{2}{3}V_{Se}^{all} \qquad (4)$$

where all quantities are in densities per square nm. Note the $Se_i$(5.5A) defect type is given an extra weight of 0.5 relative to $Se_i$(0.5A) due to it being shared between the bottom layer of the first QL and top layer of the second QL (while $Se_i$ (0.5A) is felt only by the top layer of the first QL).

| Defect densities (x10⁻² / nm²) | | | | | | |
|---|---|---|---|---|---|---|
| | $Se_i$ (0.5A) | $Se_i$ (5.5A) | $Bi_{se}$ (all) | $V_{se}$ (all) | Total | **Excess Bi** |
| **L1** | 0.10 | 0.55 | 0.83 | 0.076 | 1.6 | 1.2 |
| **L2** | 0.17 | 0.59 | 0.57 | 0.093 | 1.4 | 0.71 |
| **L3** | 0.19 | 0.53 | 0.57 | 0.093 | 1.4 | 0.72 |
| **L4** | 0.040 | 0.51 | 0.92 | 0.040 | 1.5 | 1.4 |
| **Average** | 0.13 | 0.54 | 0.72 | 0.76 | 1.5 | 0.99 |

**Supplementary Table 1: Summary of the different defects counted from the STM images of Supp. Fig. 5, and corresponding calculations for amount of stoichiometric excess Bi of the sample.**

Approximating with ~7 unit cells per square nanometer, our results indicate an excess of ~.07% Bi, and (Bi-richness) is consistent with the types of majority defects found [23]. In spite of the Bi-rich stoichiometry, Se vacancies were rare compared with excess Se. This is likely due to interstitial Bi being highly unfavored, and excess Bi instead occupying the Se vacancies as $Bi_{se}$.

**Supplementary Note 6: Comparing the surface and bulk conduction band movements from the binning series**

To corroborate the findings of the analysis summarized in Figure 4a of the main text, a separate independent analysis method of the relative band shifts is performed. Comparing EDCs between the heavily binned spectra of Fig. 3c1 and 3c3, the energy difference is estimated as the minimum energy shift required to exactly align pixel intensity between Fig. 3c3 and a spline-interpolated map of Fig. 3c1. As was seen in Figure 4a, the energy shift along the surface band comes up larger (~15 meV) in the regions where it Is energetically closer to either the bulk conduction or bulk valence band as compared to regions in the middle of the gap where it lies the farthest away from the bulk bands (about half the energy shift seen here). We further note that this analysis approach is most reliable underneath the surface band, where EDC intensity is highly sloped and has minimal direct exposure to the bulk states. The analysis will tend to be unreliable and systematically biased where the slope goes to zero, such as at points directly intersected by the red lines.

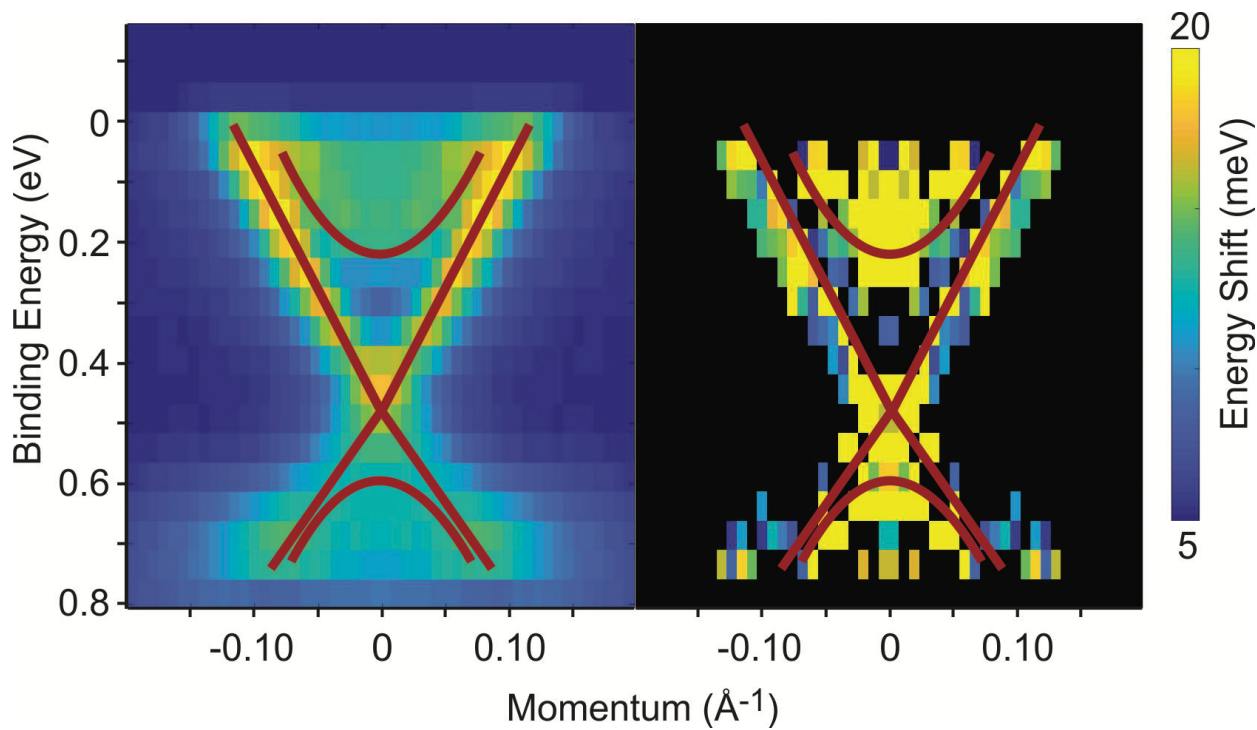

**Supplementary Figure 6: Surface band shifts across E-k space due to bulk conduction band shift. a,** An averaged spectrum including data from Fig. 3c1 and 3c3, with heavy binning to improve statistics. **b**, The energy difference between Fig. 3c1 (bulk binding energy $E_B$=0.227 eV) and 3c3 (bulk binding energy $E_B$=0.247 eV) is evaluated at each pixel with sufficient intensity to yield useful statistics. Energy difference is estimated as the minimum energy shift required to exactly align pixel intensity between Fig. 3c3 and a spline-interpolated map of Fig. 3c1. Rough guides to the eye are provided for the surface and bulk band dispersions, and the accuracy of the analysis approach varies greatly in different spectral regions as discussed in the text.

**Supplementary Note 7: Tight binding model of the Bi$_2$Se$_3$ surface**

The tight binding model of the Bi$_2$Se$_3$ surface is presented on a basis of 4 states per unit cell, as in Ref. [22], with orbital pseudospin and real electron spin addressed by $\tau$ and $\sigma$ Pauli matrices, respectively. Kinetics were set to approximately match the bulk dispersion of Cu$_{0.12}$Bi$_2$Se$_3$ [22]. A 300 meV inverted mass gap was created using the H$_0$ Hamiltonian term and set to de-invert at k=0.1Å$^{-1}$ [H$_0$=-$\tau_x$ $\Delta$/2, where $\Delta$=0.3 eV × (1-(|$\mathbf{k}_{xy}$|/0.1 Å$^{-1}$)$^2$)]. Translational symmetry within the [111] surface plane is used to define the 2D momentum vector $\mathbf{k}_{xy}$ and set the bulk Dirac velocity to 3 eV-Å (H$_D$=3 eV-Å $\mathbf{k}_{xy}$ × $\boldsymbol{\sigma}$), and particle-hole symmetry was broken with an upward-parabolic term (H$_{parab}$=10 eV/Å$_2$ • k$^2$). Spin-conserving hopping along the surface-normal axis was allowed between nearest neighbor orbitals at the top and bottom of consecutive quintuple layers (H$_t$=0.4 eV • $\mathbf{t}$). The combined Hamiltonian (H($\mathbf{k}$) = H$_0$ + H$_D$ + H$_{parab}$) was solved by exact diagonalization, with a bulk-like system size of 300 quintuple layers selected for Fig. 4c of the main text.

The following additional procedures were carried out to generate the trend curve in Fig. 4c:

1. The bulk binding energy (x-axis) was varied by applying a scalar potential to all orbitals beneath the top quintuple layer. The most intense LDOS feature in the lower half of the plot was tracked to obtain the Dirac point binding energy (y-axis). This feature represents the surface state Dirac point when bulk binding energy is large, and a bulk-degenerate resonance state when bulk binding energy is small. A kink-like feature appears at the crossover between these regimes. Aligning the kink with the experimental data places it at a point where the bulk conduction band intensity onset occurs at E$_B$~0.18 eV binding energy, and the bulk conduction band minimum is likely ~50 meV higher at ~0.13 eV. The measured Dirac point binding energy is E$_S$~0.4 eV, which should coincide with the energy of the bulk valence band maximum. Though the bulk valence band maximum of Bi$_2$Se$_3$ is not directly visible in ARPES measurements, these bulk energies are consistent within error with the ~0.3 eV bulk band gap expected for Bi$_2$Se$_3$.

2. Both axes were multiplied by 3/2 to coarsely offset the bias from sampling within a peaked distribution, as motivated by our assessment in Supplementary note 4. Sampling bias is not strictly controlled for Fig. 4c, because we do not adequately understand the tails of the measurement error distribution.

3. A constant slope was added to roughly align with the data. This correction is necessary because we have limited understanding of the actual surface environment, and have not attempted to accurately model the spatial distribution of the bulk potential. Localizing the bulk potential (tracked on the x-axis) deeper inside the crystal or extending it closer to the surface results in a shallower or steeper positive slope in the data trend, respectively.

## Supplementary References